\documentclass[atmp,nocrop,noinfo]{ipart_v1}
\Year{2026}
\firstpage{1}
\renewcommand{\jourinf}{Calabi-Yau at 50: Legacy and Future proceedings}

\renewcommand{\copyrightnotice}{%
  \vbox to\headheight{\vfil\fontsize{8}{9.5}\selectfont
    \hbox{ADV.~THEOR.~MATH.~PHYS.}\hbox{\jourinf}}}

\usepackage[T1]{fontenc}
\usepackage[utf8]{inputenc}
\usepackage[english]{babel}
\usepackage{lmodern}

\usepackage{mathtools}

\hypersetup{hidelinks,
  pdftitle={Explicit bounds in elliptic Calabi-Yau threefolds},
 pdfauthor={Seung-Joo Lee}}
\numberwithin{equation}{section}
\theoremstyle{plain}
\newtheorem{theorem}{Theorem}[section]
\theoremstyle{remark}

\newcommand{\PP}{\mathbb{P}}
\newcommand{\FF}{\mathbb{F}}

\newcommand{\QQ}{\mathbb{Q}}
\newcommand{\cO}{\mathcal{O}}
\newcommand{\cN}{\mathcal{N}}

\newcommand{\beq}{\begin{equation}}
\newcommand{\eeq}{\end{equation}}
\DeclareMathOperator{\rk}{rk}
\DeclareMathOperator{\MW}{MW}

\DeclareMathOperator{\mult}{mult}

\newcommand{\arxiv}[1]{\href{https://arxiv.org/abs/#1}{\texttt{arXiv:#1}}}

\renewcommand{\clearemptydoublepage}{\clearpage}

\begin{document}

\title[Explicit spectrum bounds in F-theory]{Explicit bounds in elliptic \\ 
Calabi-Yau threefolds:\\
dialogues between physics and geometry}

\author[Lee]{Seung-Joo Lee}
\dedicatory{Based on a talk at ``Calabi-Yau at 50: Legacy and Future,''\\ 
Tsinghua University and BIMSA, 8 September 2026.}

\begin{abstract}
We review explicit bounds on the tensor and the abelian vector sectors
of six-dimensional $\cN=(1,0)$ supergravity, with an emphasis on their
geometric derivation in F-theory and their physical interpretation
in the effective-theoretic setting.
For the tensor count, a uniform log-canonical singularity condition
on a naturally associated surface pair constrains the number of
admissible blowups, yielding the upper bound of 567.
For the abelian vector count, the nested structure of elliptic and
K3 fibrations leads to the bound of 18 for rationally fibered bases,
while a closely related surface-restriction argument reproduces the
bound of 28 for the exceptional base $\PP^2$.
On the physical side, we revisit the worldsheet unitarity argument for
suitable BPS strings and explain how the bottom-up bound of $20$ applies
also for $T\geq9$, including abelian directions without charged matter,
under the stated probe assumptions.
We highlight the common role of rational fibrations and heterotic
strings in making boundedness explicit.

\end{abstract}
\maketitle

\section{Introduction}
\label{sec:intro}

Which low-energy quantum field theories can consistently incorporate quantum gravity?
The distinction between the landscape and the swampland turns this broad
question into the search for general consistency constraints, applicable to every effective field theory
\cite{VafaSwampland}. 
While quantum field theory alone does not serve as a satisfactory UV-complete framework for describing quantum gravity, string compactifications provide a particularly useful and practical setting in which to investigate such constraints. Specifically, a geometric restriction, if established uniformly for an appropriately defined class of varieties, strongly suggests a physical consistency condition. Moreover, a physical expectation can generate a mathematical conjecture that is not apparent from geometry alone.


An important class of questions concerns the massless spectrum.
For a fixed spacetime dimension and a fixed amount of supersymmetry, can the number
of particle species, dubbed in this article simply as the particle spectrum, grow without bound? If not, can one calculate explicit
upper bounds, rather than merely addressing the boundedness? The second
question is more involved than the first and at the same time, if properly addressed, leads to more practical and exciting applications. 
An explicit bound gives a quantitative
test for a proposed effective theory and may also expose the precise mechanism that
prevents the spectrum from growing beyond the bound.

Six-dimensional $\cN=(1,0)$ theories are a useful arena for this problem.
Their chiral spectra obey restrictive anomaly equations, and a vast class
of their string realizations is described by F-theory on an elliptically
fibered Calabi-Yau threefold, 
\beq
\pi: Y \to Z\,,
\eeq
where $Z$ is the base surface. 
In this setting, the number of tensor
multiplets and the number of abelian vector multiplets, denoted respectively by $T$ and $N$, are determined by the geometry of the fibration as
\begin{equation}
\label{eq:main-dictionary}
 T=\rho(Z)-1,\qquad N=\rk\MW(\pi),
\end{equation}
where $\rho(Z)$ is the Picard number of the base $Z$ and $\rm MW(\pi)$ is the Mordell-Weil group 
of the rational sections to the fibration $\pi$. These relations translate the physical spectrum
bounds into the geometric constraints on the base Picard lattice and the rational sections.

Abstract boundedness results for elliptic Calabi-Yau threefolds
\cite{Gross,FHS} provide an important background to this discussion, but do
not by themselves supply the explicit bounds of interest here. We focus
instead on two complementary problems of establishing explicit bounds beyond boundedness. 
The first problem concerns bounding $T$ by controlling the birational
complexity of a surface pair associated with $\pi: Y\to Z$~\cite{BLPhysics,BLMath}. 
The second, on the other hand, is about bounding $N$ by restricting
the elliptic fibration $\pi$ to a suitable K3 surface embedded in $Y$,
in a way that does not miss any independent generators of the 
free Mordell-Weil group~\cite{LW,LO};  
a general mathematical treatment of the relevant
Mordell-Weil restriction arguments has subsequently been given in~\cite{GMPSW}. 

In bounding the tensor spectrum $T$ and the abelian vector spectrum $N$, various explicit bounds arise under different hypotheses via different approaches. Therefore, the resulting bounds should not be conflated: 
\begin{itemize}
\item 
First, for the tensor sector, 
$
T\leq567 
$
has been rigorously established geometrically as a top-down bound~\cite{BLPhysics,BLMath}. However, this is a rather conservative bound and efforts to improve it down to 
$
T\leq 193
$ 
as conjectured in~\cite{MorrisonTaylor2} and proposed via bottom-up arguments in~\cite{KVX}, are currently underway. 
\item
Next, for the abelian vector sector, 
$
N\leq18
$
has been geometrically derived as a top-down bound~\cite{LO} via a rational ruling of the base, while bottom-up arguments~\cite{LW} lead to 
$
N\leq20
$, 
via the heterotic-like string detecting all the abelian currents, if present. Fully general versions require further
qualifications. 
\end{itemize}
Clarifying the underlying hypotheses and the resulting bounds is part of the purpose of
this exposition.
In revisiting the earlier arguments for the abelian vector sector, we discuss two further points. 
On the top-down side, the surface-restriction argument can also be applied
to the exceptional base $\PP^2$, reproducing the bound $N\leq28$ recently established in~\cite{GMPSW}.
On the bottom-up side, we explain how the mixed gauge-gravitational
anomaly equation sharpens the bound of $22$ obtained in~\cite{LW}
for $T\geq9$ to $20$, including abelian directions without charged
matter. We make explicit why the argument establishes positive
definiteness of the full abelian level matrix under the stated
string-probe assumptions.

The article is organized as follows. Section~\ref{sec:setup} introduces the
F-theory dictionary, with special emphasis on the rational fibration structure of the base surface $Z$, 
which plays a central role in addressing both problems. 
Section~\ref{sec:tensors} explains how the tensor spectrum $T$ is bounded, 
illustrating explicit top-down computations. Sections~\ref{sec:mw} and~\ref{sec:strings} 
serve as an overview of how the abelian vector spectrum $N$ is bounded, 
via top-down and bottom-up routes, respectively. We conclude in Section~\ref{sec:outlook}
with a comparison of the mechanisms and of their limitations. 

This article is intended primarily as an exposition, with some
refinements, of the results of the author and his collaborators,
rather than a replacement for their full proofs or a comprehensive
survey of the relevant developments.

\section{Six-dimensional F-theory and the common geometric structure}
\label{sec:setup}

\subsection{The spectrum and its geometric dictionary}

We consider conventional F-theory compactifications on a projective
elliptic Calabi-Yau threefold with a rational section\footnote{For a genus-one fibration without a section, we consider
its Jacobian fibration.}
\cite{VafaFTheory}. For the purposes of establishing our bounds, we may assume without loss of generality that the base surface $Z$ is smooth.
\footnote{For the tensor bound, we use the minimal resolution of the
log Calabi--Yau pair~\cite{BLMath}. For the abelian bound, a suitable
birational elliptic model has smooth base~\cite{GMPSW},
with unchanged Mordell--Weil group.}
We further assume that the associated Weierstrass model has no non-minimal loci, either in complex codimension one or two in the base. 
We exclude frozen backgrounds and compactifications with enhanced supersymmetry.
For a general introduction to the F-theory dictionary,
we refer the reader to~\cite{WeigandTASI}.

Besides the gravity multiplet, the massless spectrum consists of tensor,
vector and hypermultiplets, with respective counts $T$, $V$ and $H$. If
\begin{equation}
 \mathfrak g=\bigoplus_\kappa\mathfrak g_\kappa
       \oplus\bigoplus_{i=1}^N \mathfrak u(1)_i,
\end{equation}
then $V=\sum_\kappa\dim\mathfrak g_\kappa+N$, and 
the cancellation of gravitational anomaly requires~\cite{KMT}
\begin{equation}
\label{eq:grav-anomaly}
 H-V+29T=273.
\end{equation}
This relation does not bound $T$ by itself: additional tensors can in principle
be accompanied by changes in the other sectors, notably by the increase of the vector spectrum $V$. 

In F-theory, the tensor charges lie in a lattice $\Gamma$ with bilinear form of signature
$(1,T)$, whose free part is identified with the intersection
lattice of $Z$. The gravitational anomaly coefficient $a$ with  
\begin{equation}
\label{eq:a-convention}
a\cdot a=9-T \,,  
\end{equation} 
is determined as the canonical class of the base $Z$,
\beq
a=K_Z \,,
\eeq
and the normalized tensor scalars
described by a vector $j$ with $j\cdot j=1$ in the appropriate positive
cone correspond to the normalized K\"ahler class.

On a smooth base $Z$ in this setting, the Picard number $\rho(Z)$ is the same as
$h^{1,1}(Z)$. One combination of the two-form fields belongs to the
gravity multiplet; this accounts for the subtraction of one in
\eqref{eq:main-dictionary}. The free part of the Mordell--Weil group, on the
other hand, produces the massless abelian vector fields
\cite{MorrisonPark}. Under the usual flat-fibration and resolution
hypotheses, the Shioda-Tate-Wazir divisor count takes the form \cite{Wazir}
\begin{equation}
\label{eq:stw}
 h^{1,1}(Y)=\rho(Z)+1+N+\sum_\kappa\rk\mathfrak g_\kappa.
\end{equation}
The extra one is the zero section, and the last term counts the resolution
divisors of the non-abelian fibers in codimension-one loci. Note that a bound on $N$ does not bound the full count $V$ of vector multiplets, nor does it alone bound
$h^{1,1}(Y)$.

The Weierstrass description is
\begin{equation}
\label{eq:weierstrass}
 y^2=x^3+Fx+G \,,
\end{equation}
where $F$ and $G$ are sections,
\beq
 F\in H^0(Z,\cO(-4K_Z)),\quad G\in H^0(Z,\cO(-6K_Z)),
 \eeq
with discriminant $\Delta=4F^3+27G^2$. The divisor of $\Delta$ lies in
$|-12K_Z|$ and records the singular elliptic fibers. As we will discuss later, this Weierstrass discriminant 
has an intimate connection to the discriminant
part of the canonical bundle formula.

\subsection{Rational fibrations, K3 surfaces and strings}
\label{sec:ruling}

The classification of smooth bases relevant here reduces the main problem
to rational surfaces, with the Enriques case treated separately. A rational base is either $\PP^2$ or can be
obtained by blowups of a Hirzebruch surface $\FF_n$~\cite{Grassi}. In the latter case it
inherits a morphism
\begin{equation}
\label{eq:ruling}
 p:Z\longrightarrow\PP^1_{\rm b}
\end{equation}
whose generic fiber $f$ is a smooth rational curve. Blowups may make special
fibers reducible or non-reduced, but do not change the generic relations
\begin{equation}
\label{eq:fiber-intersections}
 f^2=0,\qquad K_Z\cdot f=-2.
\end{equation}
Thus the absence of a rational ruling on $\PP^2$ is a genuine exception, not
an insignificant notational distinction.

The composite fibration $p\circ\pi:Y\to\PP^1_{\rm b}$ has an elliptic K3 surface $S$ as
its generic fiber. Indeed, restriction of \eqref{eq:weierstrass} to $f$
gives Weierstrass coefficients of degrees $8$ and $12$ on $\PP^1$. 
We thus obtain the nested structure

\beq\label{eq:nested}
\begin{array}{ccc}
S & \hookrightarrow & Y~ \\
\mathllap{\pi_S\;}\downarrow & & \downarrow\mathrlap{\;\pi} \\
f & \hookrightarrow & Z~ \\
 & & \downarrow\mathrlap{\;p} \\
 & & \mathbb{P}^1_{\rm b}
\end{array}
\eeq
It is this surface $S$ that will control the abelian vector count. 

The ruling $p$ has a particularly nice physical interpretation: the D3-brane wrapped on $f$
gives a distinguished string in six dimensions. Its tension is proportional to the
volume of $f$, and its worldsheet anomaly data are those of a critical
heterotic string. In a suitable infinite-distance limit the string can become
asymptotically weakly coupled and tensionless in Planck
units~\cite{LLW}, providing a concrete realization of the
Emergent String Conjecture~\cite{ESC}.
For the bounds discussed below, however, it is its existence and
its anomaly data that matter for us in bounding the spectrum; we do not consider any infinite-distance limits. This provides
a first indication of why the rational fibration structure found in rationally fibered bases $Z$ can be useful both
for geometry and for a low-energy consistency argument.

\section{Bounding the tensor sector}
\label{sec:tensors}

\subsection{From an elliptic threefold to a log Calabi-Yau pair}

The central geometric input of \cite{BLPhysics,BLMath} is more precise than
the effectiveness of $-K_Z$. The canonical bundle formula associates to
$\pi:Y\to Z$ a discriminant divisor $\Lambda_Z$ and a moduli part $M_Z$,
schematically satisfying
\begin{equation}
\label{eq:cbf}
 K_Y\sim_{\QQ}\pi^*(K_Z+\Lambda_Z+M_Z).
\end{equation}
The coefficient of a prime divisor $D\subset Z$ in $\Lambda_Z$ is
$1-t_D$, where $t_D$ is the log canonical threshold of $\pi^*D$ over the
generic point of $D$. For an elliptic fibration with section and canonical
singularities, the Kodaira fiber types give the values
\begin{equation}
\label{eq:kodaira-coefficients}
 \operatorname{coeff}(\Lambda_Z)\in
 \left\{0,\frac16,\frac14,\frac13,\frac12,
                    \frac23,\frac34,\frac56\right\}.
\end{equation}
The value $5/6$, corresponding to type $II^*$, is the largest in this list.

The reduction to a surface pair includes a suitable effective choice of
the moduli part on an appropriate birational model. More precisely, the
construction in \cite{BLMath} yields an effective boundary $B$ on the base
such that
\begin{equation}
\label{eq:pair-data}
 12(K_Z+B)\sim0,\qquad (Z,B)\text{ is }\frac16\text{-log canonical}.
\end{equation}
One way to formulate the construction is to work on a sufficiently high
resolution
\beq
\bar Z \to Z \,,
\eeq
 where $12M_{\bar Z}$ is free, take $1/12$ of a general effective representative of that
linear system, and push the resulting boundary down. The choice of
representative matters: the statement is not that an arbitrary effective
divisor linearly equivalent to the moduli part has the required
singularities.

For clarity, a pair is $\epsilon$-log canonical, or $\epsilon$-lc, if all
its log discrepancies are at least $\epsilon$. Equivalently, on any
birational model $\varphi:Z'\to Z$, the crepant pullback defined by
\begin{equation}
 K_{Z'}+B_{Z'}=\varphi^*(K_Z+B)
\end{equation}
has all boundary coefficients at most $1-\epsilon$. The condition therefore
controls exceptional divisors as well as the divisors visible on $Z$.
Checking only that the coefficients of $B$ are at most $5/6$ would not be
sufficient.

Nor should one identify $B$ literally with $\frac1{12} \Delta$. For example, an $I_n$ fiber contributes zero to $\Lambda_Z$,
although its Weierstrass discriminant has vanishing order $n$. The moduli
part and the threshold definition of the discriminant are essential to
\eqref{eq:pair-data}.

The theorem underlying the tensor bound can now be stated independently
of F-theory \cite{BLMath}.
\begin{theorem}
\label{thm:picard}
Let $(Z,B)$ be a projective $\frac16$-lc pair of dimension two, with $Z$
rational and $12(K_Z+B)\sim0$. Then
\begin{equation}
 \rho(Z)\leq568.
\end{equation}
\end{theorem}
Together with \eqref{eq:main-dictionary}, this gives $T\leq567$ for the
rational bases under discussion. The exceptional Enriques base has
$\rho(Z)=10$ and does not affect the bound. 

\subsection{Crepant blowups and a finite horizontal budget}
\label{sec:horizontal}

We outline why the hypotheses of Theorem~\ref{thm:picard} give an explicit
number. After the smooth reduction, contract the rational surface to a
Hirzebruch surface $Z_0=\FF_n$; a surface obtained by blowing up $\PP^2$
at least once can also be treated this way. The case $Z=\PP^2$ itself has
Picard number one and needs no estimate.

Starting with the induced pair $(Z_0, B_0)$, consider the sequence of point blowups
\begin{equation}
\label{eq:blowup-sequence}
 Z_r\longrightarrow Z_{r-1}\longrightarrow\cdots
       \longrightarrow Z_1\longrightarrow Z_0\,.
\end{equation}
The bounding construction chooses a maximal admissible sequence whose
endpoint dominates the original surface. Consequently,
\begin{equation}
\label{eq:r-to-T}
 \rho(Z)\leq r+2\,,\qquad T\leq r+1\,.
\end{equation}
Each step has an effective boundary $B_i$ and preserves both the crepant
relation and the $1/6$-lc condition. For a blowup at $z_i\in Z_i$ with
exceptional curve $E_{i+1}$, the coefficient of $E_{i+1}$ is
\begin{equation}
\label{eq:exceptional-coefficient}
 b_{E_{i+1}}=\mult_{z_i}B_i-1\,.
\end{equation}
Thus effectiveness demands enough boundary multiplicity at the center,
while the $1/6$-lc condition bounds the coefficient of every exceptional divisor introduced.
These are competing restrictions, not an arbitrary freedom to blow up.

Every $Z_i$ retains a rational fibration over $\PP^1_{\rm b}$. Split
\begin{equation}
 B_i=B_i^{\rm h}+B_i^{\rm v}
\end{equation}
into the components horizontal and vertical with respect to that fibration,
and associate to each blowup its horizontal multiplicity
\begin{equation}
 h_i=\mult_{z_i}B_i^{\rm h}\,.
\end{equation}
Since $12B_i$ is integral on the smooth models,
\begin{equation}
\label{eq:h-quantization}
 h_i>0\quad\Rightarrow\quad h_i\geq\frac1{12}\,.
\end{equation}
The horizontal part changes by strict transform, so the blowup formula
immediately gives the telescoping relation
\begin{equation}
\label{eq:telescoping}
 -K_{Z_{i+1}}\cdot B_{i+1}^{\rm h}
   =-K_{Z_i}\cdot B_i^{\rm h}-h_i\,.
\end{equation}
To bound the total expenditure in horizontal multiplicity, it remains to
bound the two ends of this expression.

The starting surface satisfies
$-K_{Z_0}\cdot B_0^{\rm h}\leq K_{Z_0}^2=8$, since its vertical boundary
is nef. At the other end, the $1/6$-lc condition implies a uniform lower
bound on the self-intersections of negative curves. To see the mechanism,
write $B=bC+D$, with $C$ a prime curve and $C\not\subset\operatorname{Supp}D$.
Adjunction and $K_Z+B\equiv0$ give
\begin{equation}
\label{eq:negative-curve}
 (1-b)C^2=2p_a(C)-2+D\cdot C\geq-2\,.
\end{equation}
For $C^2<0$ and $b\leq5/6$, this implies $C^2\geq-12$. In particular,
the minimal ruled surface has $n\leq12$.

On $Z_r$ one consequently has $B_r^{\rm h}\cdot C\geq-10$ for any curve
$C$. The divisor $B_r^{\rm h}+10f$ is nef: vertical curves meet the
horizontal boundary non-negatively, while a horizontal curve has
$f\cdot C\geq1$. Using $B_r\geq0$ and $-K_{Z_r}\equiv B_r$, we find
\begin{equation}
\label{eq:horizontal-endpoint}
 -K_{Z_r}\cdot B_r^{\rm h}
 =B_r\cdot(B_r^{\rm h}+10f)+10K_{Z_r}\cdot f\geq-20 \,.
\end{equation}
Summing \eqref{eq:telescoping} now yields
\begin{equation}
\label{eq:horizontal-budget}
 \sum_{i=0}^{r-1}h_i
 =-K_{Z_0}\cdot B_0^{\rm h}+K_{Z_r}\cdot B_r^{\rm h}
 \leq8+20=28\,.
\end{equation}
Hence the number $r_+$ of the blowups with a strictly positive horizontal multiplicity is subject to 
\beq
\label{eq:r-plus-bound}
r_+ \leq 28 \times 12 = 336 \,,
\eeq
where we have used~\eqref{eq:horizontal-budget} and~\eqref{eq:h-quantization}.
This is the first global constraint in the argument.

\subsection{Vertical blowups and the refinement to 567}
\label{sec:vertical}

The remaining blowups have $h_i=0$ and must be controlled differently.
The support of a fiber obtained from a ruling by point blowups has simple
normal crossings. At a smooth point of a single vertical component,
\eqref{eq:exceptional-coefficient} would give a negative coefficient,
because that component has coefficient at most $5/6$. A zero-horizontal
blowup must therefore occur at an intersection of two vertical components.

There is a useful local explanation of why even these blowups terminate.
If the two components have coefficients $b_1,b_2\leq5/6$, a divisor
obtained by successive blowups of their crossings is labelled by
a primitive pair of positive integers $(u,v)$,
i.e., with $\gcd(u,v)=1$. Its crepant boundary
coefficient is
\begin{equation}
\label{eq:local-ray}
 b_{u,v}=1-u(1-b_1)-v(1-b_2)\,.
\end{equation}
Effectiveness requires $u+v\leq6$. The local counting problem
thus reduces to counting primitive positive integer pairs
satisfying this inequality. There are just $11$ such pairs.
Thus no more than $11$ such blowups can lie over
one original crossing. This makes the numerical significance of the gap
between $5/6$ and $1$ particularly transparent.

After organizing the sequence so that the positive-$h_i$
steps come first, their number, $r_+$, bounds the number
of crossings to which this local estimate must be applied.
One thus obtains the coarse estimate
\begin{equation}
\label{eq:coarse}
 r-r_+\leq11r_+ \quad \Rightarrow \quad r \leq 12 r_+ \leq 4032 \quad \Rightarrow \quad T \leq 4033\,.
\end{equation}
Although rather conservative, 
it already exhibits a complete boundedness mechanism: 
a quantized horizontal budget limits the number of
places where additional vertical complexity can develop, and a discrepancy
gap limits that complexity locally.

The refinement uses information that \eqref{eq:coarse} discarded. We apply the
canonical bundle formula a second time, 
now to the rational fibration of the pair:
\begin{equation}
\label{eq:second-cbf}
 K_{Z_r}+B_r\equiv
 p_r^*(K_{\PP^1_{\rm b}}+\Lambda_{\PP^1_{\rm b}}+M_{\PP^1_{\rm b}})\,.
\end{equation}
Let $\mu_x^{\rm sf}$ be the coefficient of $x$ in the discriminant 
$\Lambda_{\PP^1_{\rm b}}$ on $\PP^1_{\rm b}$, 
where the superscripted $\rm sf$ indicates that a special fiber sits over $x \in \PP^1_{\rm b}$. 
Equivalently, it is one minus the log canonical threshold
of the special fiber, with respect to the pair $(Z_r,B_r)$. The moduli divisor is
nef, and the left-hand side is numerically trivial. Therefore
\begin{equation}
\label{eq:base-budget}
 \sum_x\mu_x^{\rm sf}
   \leq \deg\Lambda_{\PP^1_{\rm b}}\leq2\,.
\end{equation}
This is a second global budget, independent in character of
\eqref{eq:horizontal-budget}.

The zero-$h_i$ blowups at the blowup points on most fibers cannot saturate 
the worst-case local estimate of $11$ for their count. 
If $\mu_x^{\rm sf}<3/4$, their vertical boundary coefficients are at most
$2/3$; if $\mu_x^{\rm sf}<5/6$, they are at most $3/4$. The local counting
in \eqref{eq:local-ray} then gives $3$ and $5$, respectively, rather than
$11$. Define
\begin{align}
 X_2&=\{x\in\operatorname{Supp}\Lambda_{\PP^1_b}\,|\,3/4\leq\mu_x^{\rm sf}<5/6\}\,,\nonumber\\
 X_3&=\{x\in\operatorname{Supp}\Lambda_{\PP^1_b}\,|\,5/6\leq\mu_x^{\rm sf}\}\,.
\end{align}
By \eqref{eq:base-budget}, $|X_2|+|X_3|\leq2$: only two fibers at most
can fall into these more expensive classes.

The remaining book-keeping distinguishes blowups at a point of one fiber
component from blowups at a crossing of two. It also uses the fixed
intersection $B_r^{\rm h}\cdot f=2$ to constrain the branching of the
exceptional-curve trees in an individual fiber. Combining these local
estimates with the horizontal budget gives \cite{BLMath}
\begin{equation}
\label{eq:refined-r}
 r\leq336+66|X_2|+115|X_3|\leq566\,.
\end{equation}
The coefficients $66$ and $115$ summarize the refined fiber-by-fiber
counting; deriving them requires the additional combinatorial estimates~\cite{BLMath}. Substitution into \eqref{eq:r-to-T} thus proves
\begin{equation}
\label{eq:final-T}
 \boxed{~~\rho(Z)\leq568\,,\qquad T\leq567\,.~~}
\end{equation}

\subsection{Physical interpretation and sharpness}

Each blowup of the F-theory base increases its Picard number by one and thus
adds one tensor multiplet to the spectrum. The
geometric argument explains why the local possibility of introducing new
tensor degrees of freedom does not imply an unlimited supply in a compact
gravity model. The singularity condition bounds the complexity of local
configurations, while the horizontal and base-degree budgets constrain
how many such configurations can be assembled globally.

The $1/6$-lc property also has a physical origin: it reflects the allowed
elliptic-fiber degenerations and their associated gauge and matter
sectors \cite{BLPhysics}. It should not be replaced by a freely chosen
boundary-coefficient cutoff. Conversely, geometric restrictions on the
pair provide information beyond the single six-dimensional anomaly
relation \eqref{eq:grav-anomaly}. This is an example of compactification
geometry supplying the structure needed to turn consistency into an
explicit bound.

The number $567$ is not intended to be sharp. A well-known
example realizes $T=193$~\cite{193a, 193b} and bottom-up arguments for $T\leq 193$ have
been presented in~\cite{KVX}, using additional physical input about BPS strings,
tensor-branch boundaries, anomalies and the classification of superconformal and
little string sectors. Different approaches have different premises. The
geometric estimate \eqref{eq:final-T} does not rely on such a major classification 
nor on physical input, some of which is conjectural. 
At the same time, its derivation deliberately allows more configurations than may
actually arise from elliptic Calabi-Yau threefolds. Improving the
geometric estimate towards $\rho(Z)\leq194$ is therefore a meaningful
problem, rather than a contradiction between the resulting bounds 
from the top-down and the bottom-up approaches.

\section{Bounding the abelian vector sector geometrically}
\label{sec:mw}

\subsection{Restriction to an elliptic K3 surface}\label{subsec1}

The bounding task for the abelian vector sector is rather different 
from the tensor counterpart. 
Rational sections are global objects, and bounding their rank directly from a
Weierstrass equation is not straightforward. Interestingly, however, the nested fibration
\eqref{eq:nested} provides a way to reduce the problem 
from the threefold to the surface setting~\cite{LO}.

Let $S$ be the elliptic K3 surface over a suitable smooth fiber $f$ of the
rational ruling~\eqref{eq:ruling}. Restriction of sections gives a map from
$\MW(Y/Z)$ to $\MW(S/f)$. The important assertion is that, for an appropriate
choice of fiber, this map preserves independence in the free part:
\begin{equation}
\label{eq:mw-inclusion}
 \MW(Y/Z)\otimes\QQ\ \lhook\joinrel\longrightarrow\
 \MW(S/f)\otimes\QQ\,.
\end{equation}
This is stronger than merely saying that a section restricts to a section.
In general, specialization can create relations, and injectivity must be
justified. The height pairing provides such a justification for the
movable rational curves at hand;~\cite{GMPSW} proves restriction results
in a broader mathematical setting.

By compatibility of the Shioda map with restriction to a sufficiently
general fiber $f$, the height pairing of the restricted section $s|_S$
on the elliptic K3 surface $S$ is given by
\begin{equation}
\label{eq:height-restriction}
b_S(s|_S,s|_S)=f\cdot b(s,s)\,,
\end{equation}
where $b_S$ is the numerical height pairing associated with $\pi|_S:S\to f$,
while $b$ is the divisor-valued height pairing associated with $\pi:Y\to Z$.
For a non-torsion section $s$, a suitable positive integer
$m_s$ puts the latter in the form~\mbox{\cite{LW,LRW}}
\begin{equation}
\label{eq:height-positive}
 b(s,s)=\frac{1}{m_s^2}(-2K_Z+D_s)\,,\qquad D_s\geq0\,.
\end{equation}
Because $f$ is nef and $-K_Z\cdot f=2$, one obtains
$f\cdot b(s,s)\geq4/m_s^2>0$. A non-torsion section cannot therefore
restrict to a torsion section. Applied to all nonzero integral combinations
of independent generators of the free Mordell--Weil group, this establishes
the injectivity of~\eqref{eq:mw-inclusion}.

For an elliptic K3 surface, the Shioda-Tate relation is~\cite{SchuettShioda}
\begin{equation}
\label{eq:k3-ST}
 \rk\MW(S/f)=\rho(S)-2-\sum_v\rk R_v,
\end{equation}
where $R_v$ is the root lattice from the reducible elliptic fiber over $v$.
The two subtracted classes are the zero section and the elliptic fiber.
Since $h^{1,1}(S)=20$ and $\rho(S)\leq20$ over the complex numbers,
\eqref{eq:mw-inclusion} immediately gives
\begin{equation}
\label{eq:mw18}
\boxed{~~N=\rk\MW(Y/Z)\leq18\,.~~}
\end{equation}
This holds for the rationally fibered bases, in particular the rational
smooth bases with $T>0$ in our setup. Non-abelian fibers of $S$ can only
strengthen the surface estimate through the last term of
\eqref{eq:k3-ST}.

The salient point is the independence from the complexity of the special
fibers of $Z\to\PP^1_{\rm b}$. No matter how many base blowups are present,
the generic surface $S$ is still a K3 surface, with a fixed upper bound on
its Picard rank. The abelian vector bound is therefore uniform in $T$. This also
explains the contrast with the tensor argument: one problem counts how
complicated the degenerate fibers of the ruling can become, while the
other uses the limited divisor content of a generic K3 fiber.

\subsection{The exceptional bases and the limits of the argument}
\label{sec:P2}

The base $Z=\PP^2$, for which $T=0$, does not admit a morphism to a curve
with rational generic fiber. The K3 argument should therefore not simply
be extended to this case. Nor does a transition that preserves some
section data automatically preserve all independent generators of the
free Mordell--Weil group. This distinction is particularly relevant when
comparing rank arguments with arguments for torsion \cite{LO}.

For completeness, a closely related surface-restriction argument does
provide a numerical bound on $\PP^2$ \cite{GMPSW}. Let us restrict to a suitable
general line $\ell\subset\PP^2$ and let $S_\ell$ be the associated smooth
relatively minimal elliptic surface. Its fundamental line bundle has
degree $-K_{\PP^2}\cdot\ell=3$, so
\begin{equation}
 \chi(\cO_{S_\ell})=3\,,\qquad h^{1,1}(S_\ell)=30\,.
\end{equation}
Here the irregularity vanishes: $q(S_\ell)=h^{1,0}(S_\ell)=0$. The identities $e(S_\ell)=12\chi(\cO_{S_\ell})$
and $p_g(S_\ell)=\chi(\cO_{S_\ell})-1$ then give the second equality. 
The same height-positivity argument as in Section~\ref{subsec1} applies
to the nef curve $\ell$. Indeed, since
$-K_{\PP^2}\cdot\ell=3$, \eqref{eq:height-positive} gives
$\ell\cdot b(s,s)\geq6/m_s^2>0$ for every non-torsion
section $s$. Restriction to $S_\ell$ thus preserves
independence in the free Mordell--Weil group.
The Shioda-Tate relation then yields
\begin{equation}
\label{eq:mw28}
 N\leq\rho(S_\ell)-2\leq28\qquad \text{for}~Z=\PP^2\,.
\end{equation}
This is a useful completion of the geometric discussion, but does not
turn $18$ into a proven bound for $T=0$. The Enriques case is separate
again: in the conventional Calabi-Yau setup its Mordell--Weil free rank
vanishes \cite{LW,LO}.

Neither \eqref{eq:mw18} nor \eqref{eq:mw28} asserts that the endpoint is
realized by a Calabi-Yau threefold. An elliptic surface with a large
Mordell--Weil group need not occur as a fiber or restriction with all of
its sections extending to the threefold. Uniform upper bounds and
realizability are distinct questions.

\section{Bounding the abelian vector sector via strings and anomalies}
\label{sec:strings}

\subsection{What a string probe can measure}

The string on the rational fiber offers a physical counterpart to the
surface restriction. A BPS string with charge $Q\in\Gamma$ has a
two-dimensional $\cN=(0,4)$ worldsheet theory whose anomaly coefficients
are determined by six-dimensional inflow \cite{KSV,LW}. For the probes
of interest, removing the free center-of-mass hypermultiplet gives
\begin{equation}
\label{eq:central-charges}
 c_L=3Q\cdot Q-9Q\cdot a+2,\qquad
 c_R=3Q\cdot Q-3Q\cdot a.
\end{equation}
One must use the appropriate infrared theory and its R-symmetry in
applying these formulas; not every charge or strongly interacting string
sector is automatically an admissible probe.

The coupling to an abelian gauge field induces a worldsheet current.
Writing $b_{ij}$ for the abelian anomaly-coefficient matrix, its level
matrix is
\begin{equation}
\label{eq:level-matrix}
 k_{ij}=Q\cdot b_{ij}.
\end{equation}
In F-theory this is the intersection of $Q$ with the height pairing,
\begin{equation}
 b_{ij}=-\pi_*(\sigma(s_i)\cdot\sigma(s_j)),
\end{equation}
where $\sigma$ denotes the Shioda map. Thus a divisor pairing that detects
independent sections geometrically also measures gauge currents on the
string.

A nontrivial unitary $U(1)$ current algebra uses one unit of left-moving
central charge, independently of its positive level. If the charge-$Q$ string
detects $N_Q$ independent abelian currents, unitarity gives
\begin{equation}
\label{eq:unitarity}
 N_Q+\sum_\kappa\frac{k_\kappa\dim\mathfrak g_\kappa}
                              {k_\kappa+h_\kappa^\vee}
 \leq c_L,
\end{equation}
where the sum includes the non-abelian current algebras present on the
worldsheet. 
To bound the full abelian rank by this inequality, it is
necessary that $N_Q=N$, or equivalently that the abelian level
matrix $k_{ij}=Q\cdot b_{ij}$ have full rank.
While positivity of the diagonal entries in a fixed basis
alone would not suffice, the positivity argument of~\cite{LW}
applies equally to every nonzero integral combination of
independent Mordell-Weil generators, thereby establishing
the required full-rank property for $Q=f$.

For $Q=f$, \eqref{eq:fiber-intersections} gives
\begin{equation}
\label{eq:heterotic-central}
 Q^2=0,\qquad Q\cdot a=-2,
 \qquad (c_L,c_R)=(20,6).
\end{equation}
The full worldsheet includes the center-of-mass contribution $(4,6)$,
restoring $(24,12)$ for the critical heterotic string in light-cone gauge.
When all abelian factors are visible, \eqref{eq:unitarity} therefore implies
$N\leq20$. In the geometric setup, positivity of the restricted height
pairing is the same ingredient that prevents a non-torsion section from
becoming torsion under the relevant restriction.

The $20$ in this immediate central-charge argument should be distinguished
from the $18$ in \eqref{eq:mw18}. The latter uses the more detailed
structure of an elliptic K3 surface. A stronger bound of $16$ was argued
for in \cite{LW} by separating the worldsheet degrees of freedom carrying
the gauge currents from a neutral internal sector. As discussed in
\cite{LO}, compactification and possible gauge enhancement make the scope
of that separation a further question. Here we use $18$ as the direct
geometric conclusion, without imposing this extra condition on the
worldsheet theory.

\subsection{A bottom-up version and its assumptions}
\label{sec:bottomup}

To move beyond the territory of geometric compactifications, one can ask whether
six-dimensional consistency supplies a charge with the properties of
$f$. The analysis of \cite{LW} combines anomaly cancellation with the
existence of suitable BPS strings. Its assumptions are important: lattice
integrality and charge completeness do not, by themselves, prove that a
particular lattice vector is represented by an effective BPS string with
the required non-degenerate worldsheet theory.

As an illustration, take an odd lattice with diagonal metric of signature
$(1,T)$. Suppose that in an integral basis
\begin{equation}
\label{eq:canonical-a}
 -a=(3,-1,\ldots,-1).
\end{equation}
This is extra information about the characteristic vector, not a
consequence of $a^2=9-T$ alone. Assume further that the null charges
\begin{equation}
\label{eq:Qalpha}
 Q_\alpha=(1,0,\ldots,\underset{\alpha}{-1},\ldots,0),
       \qquad \alpha=1,\ldots,T,
\end{equation}
are represented by admissible BPS probes.
Each obeys $Q_\alpha\cdot a=-2$ and hence
has $c_L=20$. As physical string probes, they are
assumed to have positive tension, $j\cdot Q_\alpha>0$.

Write $b=b_{ii}$ for the anomaly coefficient of one abelian generator.
The gauge and mixed anomalies imply \cite{ParkTaylor}
\begin{equation}
\label{eq:abelian-anomalies}
 b\cdot b=\frac13\sum_{\mathbf q}x_{\mathbf q}q_i^4,
 \qquad -a\cdot b=\frac16\sum_{\mathbf q}x_{\mathbf q}q_i^2.
\end{equation}
Here $x_q$ includes the multiplicity and any non-abelian representation
dimension of the charged hypermultiplets. The same relations apply to any
real linear combination of abelian generators, specified by
$v=(v_1,\ldots,v_N)$, with the corresponding charges
$\sum_i v_i q_i$ and anomaly coefficient $b=\sum_{i,j}v_i v_j b_{ij}$.
Positive definiteness of the abelian gauge kinetic matrix $j\cdot b_{ij}$
requires $j\cdot b>0$ for every nonzero $v$.

The anomaly equations imply that $b\cdot b\geq0$, while positivity
of the gauge kinetic terms selects the future cone. In particular,
when $1\leq T\leq8$, $a$ is timelike, i.e., $a \cdot a = 9 - T > 0$, and 
$b$ cannot be null:
from the anomalies~\eqref{eq:abelian-anomalies} we learn that 
$b\cdot b=0$ would imply the absence of hypermultiplets
charged under the corresponding abelian generator
and hence $a\cdot b=0$, which is incompatible with a nonzero
null vector orthogonal to a timelike vector.

In fact, even for $T\geq9$, where null anomaly coefficients
are allowed, the pairing with $Q_\alpha$ must be strictly
positive. Indeed, $Q_\alpha\cdot b=0$ would require
$b=\lambda Q_\alpha$ for some $\lambda>0$, as both vectors
lie in the future causal cone and $Q_\alpha$ is null.
This would imply $a\cdot b=-2\lambda\neq0$, in contradiction
with $a\cdot b=0$ required by the anomaly equations for a
null $b$.

Since the same argument applies to every nonzero real linear
combination of abelian generators, the full abelian level
matrix is positive definite. Under the probe assumptions
above, we thus obtain
\begin{equation}
\label{eq:physical20}
 N\leq20\qquad \text{for}~T\geq1\,.
\end{equation}

\subsection{Refinement of the original probe bound and exceptional cases}

The original analysis of~\cite{LW} treated the possibility of null
abelian anomaly coefficients for $T\geq9$ by combining the
current-algebra constraints from the $T$ distinct probes
\eqref{eq:Qalpha}. This led to the more conservative bound
\begin{equation}
\label{eq:physical22}
 N\leq\left\lfloor\frac{20T}{T-1}\right\rfloor
 \leq22\qquad\text{for}~T\geq9\,.
\end{equation}
The mixed gauge-gravitational anomaly equation excludes a nonzero null
abelian anomaly coefficient parallel to a null probe with $Q\cdot a=-2$,
an observation also made in Appendix~C of~\cite{KVX}.
Section~\ref{sec:bottomup} makes explicit how this establishes positive definiteness
of the full abelian level matrix. A single such probe therefore
suffices to establish the bound of $20$, even in the presence of
abelian directions without charged matter.
In particular,
the refinement does not require any additional assumptions
on the string spectrum.

Although the above discussion was formulated for a particular
choice of charge lattice, the refined positivity argument
does not depend on this choice. The essential additional
physical input is the existence of an admissible
future-directed null probe with $Q\cdot a=-2$.
The case $T=0$, on the other hand, requires a separate
treatment~\cite{LW}, since the charge lattice admits no
nonzero null vector.
In this case, the unit probe with $a=-3$ has $c_L=32$, rather than 20.
This agrees with the need to analyze $\PP^2$ separately on the geometric
side, although the geometric estimate \eqref{eq:mw28} is stronger than
this elementary central-charge bound. 
Subsequent physical work addresses the more general problem
of bounding the full massless spectrum with additional
structural input~\cite{KVX}; our emphasis here has been on
clarifying and refining the original string-probe argument.


The point of the comparison is therefore not that all derivations give
one identical integer. Rather, geometry both guarantees the required
objects and adds information about them. On a rationally fibered F-theory
base, an effective null curve, a K3 restriction and a heterotic string are
three aspects of the same construction. In an abstract supergravity
theory, the existence and properties of their proposed counterparts must
be justified independently.

\section{Conclusions and outlook}
\label{sec:outlook}

The tensor and the abelian vector sectors illustrate two ways to promote geometric
boundedness to explicit spectrum bounds. For tensors, the essential
information is encoded in a log Calabi-Yau surface pair: the discrepancy
gap restricts local blowup configurations, while intersection numbers and
the canonical bundle formula on the ruling base provide global budgets.
For abelian vectors, independent rational sections survive restriction to
an elliptic surface whose Picard number is bounded. These mechanisms lead
to
\begin{equation}
 \begin{aligned}
  T&\leq567 &&\text{for the F-theory bases considered here},\\
  N&\leq18  &&\text{for rationally fibered bases},\\
  N&\leq28  &&\text{for the exceptional base }\PP^2.
 \end{aligned}
\end{equation}
The geometric hypotheses have been specified above and the exceptional
Enriques case does not weaken the bounds. The top-down bound for
$\PP^2$ complements the earlier discussion of rationally fibered bases
in~\cite{LO} and reproduces the bound established in~\cite{GMPSW}.
On the bottom-up side, the mixed-anomaly argument reviewed above
sharpens the original bound of $22$ in~\cite{LW} to $20$ for $T\geq9$.
Under the string-probe assumptions specified in Section~5, a single
suitable heterotic-like string suffices to establish $N\leq20$
for any $T\geq1$, irrespective of whether charged matter is present.

The shared rational fibration is particularly instructive. Its special
fibers carry the complexity responsible for extra tensors, whereas a
generic fiber selects both an elliptic K3 surface and a heterotic string
that constrain the abelian sector. In this sense, the same geometric
structure controls two quite different pieces of the low-energy theory.
The connection is useful in both directions: physical consistency
motivates which numerical questions to ask, and algebraic geometry
explains why their answers can be uniform across a large class of vacua.

Several tasks remain. The tensor bound is visibly non-optimal, and a
geometric derivation of the expected sharp bound would require more
precise use of the allowed fiber configurations. For the abelian vector sector,
the distinction between an upper bound and an attainable rank deserves
further investigation, especially for $\PP^2$. More generally, neither
bound settles the full classification of six-dimensional vacua: the
non-abelian sector, charged matter, and other discrete data require their
own constraints. What the results reviewed here provide is a concrete
set of methods, and calculable numbers, with which to pursue that larger
problem.

\section*{Acknowledgments}
The author is indebted to Caucher Birkar, Paul-Konstantin Oehlmann and Timo Weigand 
for the fruitful collaborations on which much of this article is based. 
He thanks the organizers of ``Calabi-Yau at 50: Legacy and Future''
for the opportunity to present the work reviewed here. 
This work was supported by a National Research Foundation of Korea (NRF) 
grant funded by the Korean government (MSIT) (Grant No. RS-2026-25604009) 
and by the Yonsei University Research Fund (Grant No. 2026-22-0183).


\begingroup
\raggedright

\endgroup

\address{
Department of Physics, Yonsei University\\
Seoul 03722, Republic of Korea\\
\email{seungjoolee@yonsei.ac.kr}\\
}

\end{document}